# STANDARD AND STRESSED VALUE AT RISK FORECASTING USING DYNAMIC BAYESIAN NETWORKS

Eden Gross, School of Statistics and Actuarial Science, University of the Witwatersrand, Johannesburg, South Africa. eden.gross@wits.ac.za ORCiD: 0000-0002-7647-8890

Ryan Kruger, Department of Finance and Tax, University of Cape Town, Cape Town, South Africa. ryan.kruger@uct.ac.za

Francois Toerien, Department of Finance and Tax, University of Cape Town, Cape Town, South Africa. francois.toerien@uct.ac.za ORCiD: 0000-00002-6051-1094

## ABSTRACT

This study introduces a dynamic Bayesian network (DBN) framework for forecasting value at risk (VaR) and stressed VaR (SVaR) and compares its performance to several commonly-applied models. Using daily S&P 500 index returns from 1991 to 2020, we produce 10-day 99% VaR and SVaR forecasts using a rolling period and historical returns for the traditional models, while three DBNs use both historical and forecasted returns. We evaluate the models' forecasting accuracy using standard backtests and forecasting error measures. Results show that autoregressive models deliver the most accurate VaR forecasts, while the DBNs achieve comparable performance to the historical simulation model, despite incorporating forward-looking return forecasts. For SVaR, all models produce highly conservative forecasts, with minimal breaches and limited differentiation in accuracy. While DBNs do not outperform traditional models, they demonstrate feasibility as a forward-looking approach to provide a foundation for future research on integrating causal inference into financial risk forecasting.

**Statements and Declarations:**

*Competing Interests and Funding*
The authors declare that they are not aware of any competing interests that exist with regards to this research. This research did not receive any specific funding from any funding agency in the public, commercial, or not-for-profit sectors.

*Declaration of the use of generative artificial intelligence in scientific writing*
During the preparation of this work, the author(s) used Copilot to improve the flow and grammar of the manuscript. After using this tool/service, the author(s) reviewed and edited the content as needed and take(s) full responsibility for the content of the published article.

# 1. Introduction

Systemic bank failures have repeatedly been felt throughout the global economy, underscoring the critical importance of robust market risk measurement and management, where, for banks, market risk is defined as the risk of losses in portfolio value due to market price movements for both on- and off-balance sheet accounts (Basel Committee on Banking Supervision, 2019a). Any risk that threatens banks' stability is of critical importance to regulators and financial policy makers, particularly given the destructive effects of systemic bank failures on the real economy, as evidenced by the 2008 global financial crisis. Consequently, banks are highly regulated, typically in line with the supranational regulatory framework of the Basel Committee on Banking Supervision (BCBS), known as the Basel Accords, which govern how banks quantify their market risks. While the BCBS allows banks to use internal models, such as the historical simulation model, for this purpose (Basel Committee on Banking Supervision, 1996), these models generally have a poor record when it comes to adequately reflecting the market risk banks are exposed to, especially during financial crises. This is evidenced by the 2008 global financial crisis, the collapse of Lehmann Brothers, and the failure of Silicon Valley Bank in 2023 in the United States (US), which was catalyzed by a market risk event (Mettick, 2024).

To manage their market risks, the BCBS requires banks to calculate market risk metrics. A common risk management metric employed by banks to measure market risk is value at risk (VaR), which calculates an expected loss of portfolio value over some target time horizon within a pre-determined confidence interval (Shenoy & Shenoy, 2000). It is a tail risk metric, calculated using the profit and loss probability density function (PDF). In addition, the BCBS introduced stressed VaR (SVaR) as an augmenting measure for VaR, where SVaR is calculated similarly to VaR, except that SVaR uses a stressed period, which must include the 2008 global financial crisis, as the historical period. While the BCBS has recently moved its market risk metric of choice from VaR to expected shortfall (ES), the US Securities and Exchange Commission (SEC) requires US banks to publish VaR forecasts (United States' Securities and Exchange Commission, 1997) in addition to their ES forecasts, and banks and regulators still rely on VaR forecasts (Liu & Stentoft, 2021). Such reporting accomplishes several goals. From a public information standpoint, VaR is a simple figure to quote, capturing a probability, a time horizon, and a loss figure, which makes it easy to both understand and use for comparative purposes. From a regulatory standpoint, while VaR forecasts are not necessarily the sole input in determining how much capital a bank must hold in reserves in the US, these forecasts are still monitored by the banking regulator and serve as input for evaluating the performance of the internal models used by banks via a process known as backtesting.

The inherent tension in a bank's ability to balance capital reserves against the cost of not holding sufficient reserves may lead to the mismanagement of market risk. Higher capital reserves, driven by higher VaR forecasts, can lead to lower bank profitability. Therefore, banks may be incentivized to use internal models which understate VaR, while maintaining a number of breaches which is not considered so excessive as to alarm the regulator (McAleer & da Veiga, 2008). The historical simulation model is often used as banks' internal model to this end due to its ease of implementation and low computing requirements (Pérignon & Smith, 2010). On the other hand, some studies find that VaR forecasts provided by banks, especially those forecasted using the historical simulation model, to be conservative and, therefore, inaccurate (see, for example, Berkowitz and O'Brien, 2002; Berkowitz, Christoffersen, and Pelletier, 2009; and Pérignon, Deng, and Wang, 2008).

Advances in computing power and the growing adoption of machine learning in finance present an opportunity to address these shortcomings of traditional models. By introducing machine learning models and techniques to quantify banks' market risk and, most importantly, to incorporate forward-looking predictions, we aim to revisit existing financial risk management practices using modern tools. Specifically, Shenoy and Shenoy (2000) and Demirer, Mau, and Shenoy (2006) proposed the use of Bayesian networks (BNs), a machine learning application of Bayesian inference in the form of a graphical representation of probabilistic causal relationships, for market risk calculations. More recently, Apps (2020) developed a simplified BN methodology to predict the directional move of a portfolio's returns and its impacts on VaR. Thus, BNs can be used to learn the causal relationships between various variables that influence a bank's trading desk's returns, use these to produce forward-looking return predictions, and then use the latter to produce more accurate tail risk metric forecasts. However, we are not aware of any study to date that attempts to provide a complete methodological application of BNs in market risk management. This study attempts to fill this gap using BNs and the equities trading desk of a generic bank using the Standard and Poor's (S&P) 500 index as a case study.

An evaluation of VaR and SVaR utilizing a BN would eliminate several of the key flaws of the traditional models. First, since a BN's output is a PDF, its use avoids assuming a distribution for the set of value predictive factors, which is common in many of the currently used models.

This is particularly important given the substantial evidence that financial returns are not normally distributed, as assumed by many prevailing models, instead displaying leptokurtic characteristics with significantly heavier tails. This suggests the existence of more extreme outliers than predicted under the normal distribution assumption (Peiró, 1999). Further, if a BN was applied to forecast VaR instead, the normality assumption would not be required, as the belief network would output a PDF from which a forecasted return can be determined for VaR forecasting. Second, in contrast to most traditional VaR models, the BN and the return PDF achieved by the equity portfolio allow for the calculation of the contribution of each previous return, which can be weighted according to its importance as determined by the belief network. Last, existing models, such as the RiskMetrics model, have a bad track record of adapting to new regimes and predicting crises like the 2008 global financial crisis. Hence, a dynamic model whose predictive abilities allow for the discarding of irrelevant information coupled with the efficient ability to adapt to the regime-switching nature of crises is needed. Thus, in several respects, a BN-based approach may offer more robust VaR estimates that can be updated as soon as new information is available.

While static BNs model causal relationships that exist at a specific point in time (Friedman, Murphy, & Russell, 2013), dynamic BNs (DBNs) include a temporal element that extends the network's capacity to learn the causal relationships within and between the nodes over time, not only causal relationships between nodes at a specific point in time (Dagum, Galper, & Horvitz, 1992). The temporal elements of these DBNs are much better suited to VaR and SVaR forecasting, as we use financial time series data to produce market risk metrics. Moreover, the time series data relating to a single variable are affected by that variable's filtration system up to that period as well as those of other related variables (i.e., those displaying causal relationships with the target variable). Thus, this study strictly uses DBNs in its methodology, and further references to BNs will mean DBNs, and references to any BNs learning algorithms will mean their respective dynamic extensions.

This study offers several contributions to the literature. First, we provide a comprehensive comparison of the accuracy of traditional market risk models in forecasting VaR and SVaR. This is done over an extended out-of-sample period from 1991 to 2020, covering several financial crises and volatility regimes. We offer this comparison for both VaR and SVaR, where the literature surrounding the performance of traditional models in forecasting the latter risk measure is incredibly scarce. Second, we do this using two underlying return distributions for appropriate models, being the normal and skewed Student's t distributions, recognizing that the latter's skewness and heavier tails may better capture equity return characteristics and improve forecast accuracy. We use both distributions to robustly evaluate their fits to the return data and their appropriateness in market risk metric forecasting. Third, in calculating SVaR, lacking guidance from the BCBS, we adopt what we believe to be the most appropriate and conservative approach to constructing the stressed period required for the forecasts. This, too, is a contribution to the literature. Fourth, we build on the theoretical basis presented by Shenoy and Shenoy (2000) and Demirer, et al., (2006), and the simplified methodology provided by Apps (2020), and develop a comprehensive BN methodology for modeling causal relationships between financial and economic variables. This approach enables us to produce one-day-ahead market risk forecasts using a rolling period methodology and several BN learning algorithms for each day in the out-of-sample period. While the results for the BNs do not suggest that banks should adopt them as a replacement for traditional market risk models, our aim is to provide the first practical application of this comprehensive methodology as a starting point for future research to further develop these ideas and explore whether the methodology we provide can be improved further in asset pricing, in general, and market risk management, in particular. Finally, we assess model accuracy using a range of backtesting techniques and forecasting error measures drawn from the literature.

The remainder of this paper is structured as follows. Section 2 provides a literature review and the technical context. Section 3 then discusses the data and methodology, while Section 4 discusses the results and provides analyses. Last, Section 5 concludes this article.

## 2. Literature Review and Technical Context

For a $100\alpha\%$ $h$-day VaR figure at time $t$, a number $x_{ht,\alpha}$ of the random variable $X$ representing the $\alpha$ quantile of the profit and loss account must be found such that

$$\Pr[X_{ht} < x_{ht,\alpha}] < \alpha \qquad (1)$$

This, then, results in the VaR at time $t$ being

$$VaR_{ht,\alpha} = -x_{ht,\alpha} \times P_t \qquad (2)$$

where $P_t$ represents the total portfolio value at time $t$ (Alexander, 2008).

To determine VaR using Equation (2), we assume that portfolio profit and loss amounts are either available or are estimated from an assumed underlying distribution (Pritsker, 2006). This assumption is critical for the accuracy of VaR forecasts.

In response to the 2008 global financial crisis, the BCBS has complemented the VaR forecast with a SVaR forecast, being a one-year VaR calibrated under stressed conditions (Basel Committee on Banking Supervision, 2009). The stressed period is taken to be the bank's most severe one-year period of losses available (Liu & Stentoft, 2021), while still quoted as a 10-day 99% figure. Since SVaR is equal to or greater than VaR (Liu & Stentoft, 2021), the capital held by banks has at least doubled following the introduction of SVaR and, moreover, has stabilized given that the period used to calibrate SVaR changes less frequently than that used to calculate VaR.

The lack of prescription offered by the BCBS allows for flexibility when it comes to calculating SVaR (or any other stressed metric). The BCBS states that the stress period must be at least one year in length, and the historical period used to determine the stress period must include 2007 (Basel Committee on Banking Supervision, 2023). However, the BCBS is not clear as to how this period is to be determined. This ambiguity offers an opportunity for banks and regulators to apply judgement – a judgement which may or may not be in line with the result intended by the BCBS. For example, a bank may choose to use a one-year period which includes the worst return observed in its historical period. This period may or may not be the worst stressed period in totality, as the other days included in the period may have been better in aggregate relative to another historical period which does not include that worst return observed, i.e., there may exist a period better suited as a stress period which does not include the particular day on which the worst return was observed. Moreover, it is possible for the stressed period to not consist of consecutive days at all. In the strictest and most conservative definition of a stressed period, a bank may amalgamate a collated set of days on which the worst returns were achieved in its historical period into a collated stressed period, thereby yielding the absolute worst calibration set. This, in fact, is the method adopted in this study to offer the most robust and conservative set of results.

In practice, one of the typical underlying assumptions when modeling VaR is that the statistical properties of financial data are stable throughout time (Daníelsson, 2002), even in times of a crisis. A model that does not assume the stationarity of the underlying process, i.e., one that can update its statistical relationships as time passes, is expected to forecast more accurately than models that cannot do so. The traditional models most commonly used by banks, and those we consider in this study, such as the historical simulation model, the delta-normal model, autoregressive models, such as the autoregressive conditional heteroscedasticity (with order 1) (ARCH(1), see Equation (3)), generalized ARCH(1,1) (GARCH(1,1) see Equation (4)), and exponential GARCH(1,1) (EGARCH(1,1) see Equation (5)) models, as well as the RiskMetrics model (see Equation (6)), fall into this latter category. Their defining equations are as follows.

$$\sigma_{t+1}^2 = \omega + \alpha \varepsilon_t^2, \quad \omega > 0, \quad \alpha \geq 0, \quad \varepsilon_t = \sigma_t z_t \tag{3}$$

$$\sigma_{t+1}^2 = \omega + \alpha \sigma_t^2 \varepsilon_t^2 + \beta \sigma_t^2, \quad \alpha, \beta, \omega > 0, \quad \alpha + \beta < 1 \tag{4}$$

$$\ln(\sigma_{t+1}^2) = \omega + \beta \ln(\sigma_t^2) + \alpha \left( \frac{|\varepsilon_t|}{\sigma_t} - E\left[\frac{|\varepsilon_t|}{\sigma_t}\right] \right) + \gamma \cdot \frac{\varepsilon_t}{\sigma_t} \tag{5}$$

$$\sigma_{t+1}^2 = \lambda \sigma_t^2 + (1 - \lambda) r_t^2 \tag{6}$$

where $\sigma_t$ is the return volatility; the $z_t$ is a series of independently and identically distributed standard normal random variables (Bollerslev, 2007); for the GARCH(1,1) model, the process's long-term average is $\sqrt{\omega/[1 - (\alpha + \beta)]}$; the $\varepsilon_t$ is the residual term of the time series; for the RiskMetrics model, $\lambda = 0.94$ is the smoothing parameter; and $r_t$ is the previous day's return.

McAleer and da Veiga (2008) calculate VaR forecasts for portfolios of several indices and find that banks are encouraged to calculate VaR using models which understate VaR while maintaining a number of breaches which is not considered so excessive as to alarm the regulator. This means that banks can benefit by employing VaR models that lead to multiple breaches and, therefore, a higher penalty, as the penalty incurred is still cheaper than the additional capital expected to be carried by banks at the standardized regulatory capital level required of them (a level which is often higher than that calculated by the internal model). This allows a bank to balance its opportunity costs and the cost of additional capital held due to the breaches, resulting in a cost-benefit analysis which may suggest that the bank should hold less capital and experience more breaches.

Sharma (2012) summarizes the historical simulation model by describing it as a model which backtests well and conforms to regulatory requirements, while displaying mixed performance under hypothesis testing scenarios, and failing when the independence of breaches is tested. This suggests that the historical simulation model would be the primary model for the aforementioned cost-benefit analysis exercise. In addition, Berkowitz and O'Brien (2002) find that the model produces conservative and inaccurate forecasts, while Pérignon and Smith (2010) find it to be the most common VaR forecasting model for international banks, producing forecasts that are often conservative, leading to few breaches. Moreover, Pérignon and Smith (2010) find that the historical simulation model suggests very little about future market volatility and, hence, provides VaR forecasts which are questionable and of little use. Hence, due to its widespread popularity, we use the historical simulation model as the base case for our analyses.

More recently, O'Brien and Szerszeń (2017) test the performance of risk measures of US banks before, during, and after the 2008 global financial crisis. The authors find that the internal models used by banks produced conservative VaR forecasts, leading to few breaches in the periods preceding and following the global financial crisis. On the other hand, crisis period (June 2007) breaches were significantly higher than predicted and clustered often (O'Brien & Szerszeń, 2017). The authors highlight the banks' models' inability to adapt to changes in market conditions and the slow updating given new information, specifically during periods of market turmoil. While the banks' internal models produced conservative VaR forecasts during more stable periods (those preceding and following the crisis), supporting the findings of studies such as that of Berkowitz and O'Brien (2002), autoregressive models experienced fewer breaches and increased independence among such breaches during the crisis. This shows the increased benefits of more accurate and timely volatility adjustments relative to models such as the historical simulation model.

There exists ample evidence that the returns of financial instruments are not normally distributed but are, in fact, leptokurtic, implying higher peaks and heavier tails relative to the normal distribution. This, in turn, suggests the existence of more extreme outliers than predicted under the normal distribution assumption (Peiró, 1999). The assumption is further contradicted when it comes to smaller datasets which cover a shorter period, as it cannot be assumed that the central limit theorem applies then. Since the introduction of the 1996 Market Risk Amendment (see Basel Committee on Banking Supervision, 1996), banks have been granted permission to make use of more sophisticated VaR models as their internal models, making the use of autoregressive models more common in financial applications. The heteroscedastic attribute of these autoregressive models allows them to incorporate non-constant volatility experienced in the underlying time series into their models (Giot & Laurent, 2004). This is a clear advantage of the autoregressive models when compared to less sophisticated models, such as the historical simulation model and the delta-normal model, as these are inefficient in incorporate changes in volatility in comparison.

Applying BNs to determine VaR has received little attention to date. BNs allow for an update of a belief system as new information becomes available. Since incorporating the heteroscedastic nature of volatility is believed to enhance US banks' internal models' performance when it comes to producing VaR forecasts (O'Brien & Szerszeń, 2017), we posit that the recalibration and updating methodology of BNs presented in this study will enhance the volatility updating nature of autoregressive models. Hence, we aim to provide a comprehensive methodology for forecasting market risk metrics using BNs that should offer more timely and robust volatility capturing. This methodology is based on the ideas of Shenoy and Shenoy (2000) and Demirer, et al., (2006), which suggest that utilizing BNs to supplement existing financial risk management practices could produce more accurate VaR estimates.

Due to the causal relationships depicted by BNs, they are well-suited as an aid for analysts examining both individual stocks as well as portfolios in the presence of uncertainty (Demirer, et al., 2006). This makes BNs an appropriate, useful, and relevant tool in the pricing of assets and portfolios and, by extension, banks' trading desk portfolios subject to regulation. Both Demirer, et al., (2006) and Shenoy and Shenoy (2000) provide basic outlines for constructing BNs for portfolio management.

It is, therefore, surprising that so little research on the appropriateness of BNs for VaR forecasting exists. The only prior study in this regard that we are aware of is that of Apps (2020), who applies a BN to calculate daily VaR forecasts relating to three United Kingdom bank shares. Apps (2020) uses a simplified methodology to model stock returns by modeling whether the three-share portfolio returns are positive or negative using a Gaussian BN[1]. The network constructed includes three variables – a liquidity variable, a market variable, and the target variable, being whether the return achieved on the three-share portfolio is positive or negative.

Hence, this study aims to fill several gaps in the literature. First, by forecasting 10-day 99% VaR and SVaR for the equities trading desk of a US bank using traditional models over a long out-of-sample period, we assess the forecasting accuracy of these models and contribute to the scarce literature surrounding SVaR. Second, we assess model forecasts when calibrated using either the normal or skewed Student's t distribution, recognizing the latter's better theoretical fit to equities return due to heavier tails and higher skewness relative to the former. We do this to robustly evaluate the traditional models further. Third, we further add to the SVaR literature by utilizing a conservative approach to stressed period construction, to enhance the reliability of the SVaR models' results. Last, we develop the ideas presented by Shenoy and Shenoy (2000) and Demirer, et al., (2006), by using several network learning algorithms to produce forecasts of both market risk metrics, as would be required by practitioners in charge of the market risk management function of the equities trading desk of a US bank.

---

[1] A Gaussian BN assumes multivariate normality (Grzegorczyk, 2010).

## 3. Data and Methodology

Daily closing values of the S&P 500 index covering 15 March 1991 to 14 February 2020 were obtained from the Bloomberg database. This period covers approximately three business cycles (National Bureau of Economic Research, n.d.). The BN model requires a large period to facilitate training and the implementation of a rolling period methodology, making this long period appropriate. The S&P 500 index serves as a proxy for the equities trading desk of a US bank, as it is representative of the US equities market (see, for example, Grinold, 1992) and mimics large capitalization stocks, which represents most of the equity exposure of banks and the most liquid stocks. The daily closing levels of the S&P 500 index were collected for the entire period, as well as for two sequential five-year periods preceding the entire period (a training period and a calibration period), each being 1,264 days long, or approximately five trading years. The later calibration period was used to calibrate the models and produce the two market risk forecasts, as is usual in the literature, while the earlier training period was used to train the BNs. We calculated daily returns for the index as proxies for the bank's daily profit and loss amounts to produce the 10-day 99% S/VaR forecasts.

The daily returns earned on the S&P 500 index were calculated using the following formula.

$$r_t = \ln\left(\frac{P_t}{P_{t-1}}\right) \tag{7}$$

where $r_t$ is the daily return earned on the S&P 500 index at time $t$; $P_t$ is the level of the S&P 500 index at time $t$; and $P_{t-1}$ is the level of the S&P 500 index on the previous trading day. In total, 7,286 daily returns were calculated in the out-of-sample period, with an average daily return of 0.03% and a standard deviation of 1.10%. The minimum daily return experienced was -9.47%, while the maximum daily return experienced was 10.96%. The returns distribution had a skewness coefficient of -0.28 and a kurtosis of 12.10.

As the BCBS requires banks to calculate 10-day 99% S/VaR forecasts, and following the guidelines of the BCBS, we scaled the one-day 99% S/VaR forecasts to 10-day 99% S/VaR forecasts using the square-root-of-time rule. For SVaR, we used a stressed period, which is the most severe period preceding the return's date and corresponds in length to the non-stressed calibration period. The BCBS states that the stress period must be at least one year in length, and the historical period used to calibrate the stress period must include 2007 (Basel Committee on Banking Supervision, 2023). As discussed previously, the BCBS is not clear as to how this period is to be determined, and it is possible for the stressed period to not be constructed of consecutive days at all. To enhance the robustness of forecasts, we adopt the strictest and most conservative definition of a stressed period by amalgamating a collated set of days on which the worst returns were achieved in its historical period into a collated stressed period.

The traditional models evaluated in this study are the historical simulation, delta-normal, ARCH(1), GARCH(1,1), EGARCH(1,1) and RiskMetrics models. For these models, we used the daily returns and a five-year rolling calibration period to calculate 7,286 out-of-sample 10-day 99% S/VaR forecasts directly. Following many other studies, the autoregressive and RiskMetrics models were calibrated using log-likelihood maximization and the normal distribution as the underlying return distribution, and the RiskMetrics parameter value for $\lambda$ was the default value of 0.94. In addition, since equity returns are often found to not be normally distributed, we also calibrate these models using the skewed Student's t distribution, which we believe fits equity market the data better and is also used by some other studies when modeling tail risk metrics[2].

For the BN models, three network learning algorithms were first trained using a rolling period methodology starting with the initial five-year training period. Using this period, each BN produced a one-day-ahead closing value for the S&P 500 index for each day of the five-year calibration period and the 7,286-day out-of-sample period. The availability of data ten trading years prior to the start of the out-of-sample period (i.e., 15 March 1991) influenced the availability of economic and financial variables' data serving as nodes in the network. The rolling period re-training methodology took place both at a point in time and between periods, utilizing the temporal element of the learning algorithms used. We deem this approach to be the most robust approach, as it allowed for new market, economic, and financial information to be incorporated into the prediction of the following trading day's closing value of the S&P 500 index, as would be the case in practice. The one-day-ahead forecasts of the closing values of the S&P 500 index were then used to calculate one-day-ahead daily returns for the index using Equation (7). These were incorporated into the return PDF to facilitate the calculation of 7,286 out-of-sample 10-day 99% S/VaR forecasts.

---

[2] See, for example, Giot and Laurent (2003); Nieto and Ruiz (2016); and Lambert and Laurent (2001).

As nodes, each of our BNs used 41 macroeconomic and financial variables, or their appropriate proxies (see Appendix A), which were either indicated by the literature or by expert judgement as having a potential causal relationship with the S&P 500 index. Depending on the nature of a specific variable, its updating frequency ranged from daily to quarterly. Since daily closing values of the S&P 500 index were forecasted by the BNs, and due to the inability of the network learning algorithms employed to handle missing values, any non-daily variable data were transformed to daily data by setting the variable to its last known value. We adopted this method as it mimics the use of available data in practice (e.g., only the last known US consumer price index value is available to the risk manager at a US bank prior to its latest value being released).

Five of the 41 variables used as inputs (daily prices of the West Texas Intermediate, the closing values of the Financial Times Stock Exchange 100 index, the daily three-month US London inter-bank offer rate values, the daily spread on ten-year Aaa-rated US corporate bonds relative to Treasury bills, and the US policy uncertainty index) did not have data for the entirety of the initial training period. Due to the learning algorithms' inability to handle missing data, these variables' earliest known values were filled in from their earliest occurrence back to the beginning of the initial training period. At most, this amounted to 981 trading days, or 10% of the total period. However, it is worth stressing that we use a rolling period methodology to re-train the BN learning algorithms at every iteration throughout the out-of-sample period. Therefore, while these filled-in values exist strictly in a portion of the initial training period employed, i.e., at most from day 0 to day 981, these values were filtered out by the time the rolling period methodology began producing forecasts beyond the initial calibration period, thereby learning the true causal relationships that existed between these variables, other variables, and the closing values of the S&P 500 index to produce one-day-ahead forecasts of the latter.

Operationalizing BN-based S/VaR estimates within a machine learning context requires a specific algorithm to learn both the structure of the network as well as its parameters (the causal relationships between the nodes). The learning algorithms used for the BN part of this study were the Peter and Clark (Stable) (PC (Stable)) algorithm (Spirtes & Glymour, 1991), the max-min hill-climbing (MMHC) algorithm (Tsamardinos, Brown, & Aliferis, 2006), and the semi-interleaved HITON parents and children (SI-HITON-PC) algorithm (Aliferis, Tsamardinos, & Statnikov, 2003). Thus, we ran three BN models for each of the S/VaR components of the study. These were applied using the rolling period methodology previously described. The network's structure was determined by minimizing the Akaike information criterion (AIC).

The approach outlined above is essentially the top-down approach outlined by Demirer, et al., (2006). The BN construction and the network learning were performed in R, using, amongst other resources, the dbnR package for the training of the DBNs employed in this study and forecasting of the one-day-ahead forecasts of the closing values of the S&P 500 index.

The implementation of the BN model to calculate the market risk metrics at the trading desk level allows for a like-for-like comparison with existing traditional S/VaR models. Moreover, restricting the scope to the equities trading desk allows for the simplification of the BN models employed. This, in turn, makes the BN models outlined in this study significantly more viable for implementation by banks. Further, the separation allows for the measurement of the suitability and robustness of BN models to different trading desks in future research. This allows for conclusions to be made on the appropriateness of BN models for the equities trading desk, as well as others.

Using the 10-day 99% VaR and SVaR forecasts generated by the traditional models and the three BNs and ten traditional models, we backtested the models, in line with the requirements of the BCBS guidelines (Basel Committee on Banking Supervision, 2019b). This process involves calculating the number of breaches experienced by each model and then statistically testing the accuracy of the model used to produce these breaches using an appropriate backtest. A breach is recorded when the daily loss made on the S&P 500 index exceeded the daily 10-day 99% S/VaR forecast. Since the historical simulation model is the most popular model employed by banks to forecast VaR (Pérignon & Smith, 2010), we use it as a base case. Once we calculated the number of breaches for each model, we backtested the various 10-day 99% S/VaR forecasts using the BCBS's traffic light test, which is the BCBS's regulatory VaR forecasting model backtest, as well as the Kupiec's proportion of failure coverage test (Kupiec, 1995) and Christoffersen's test for independence (Christoffersen, 1998), which tests for a clustering of breaches.

While backtesting is a requirement and a standard procedure when evaluating the statistical accuracy and suitability of the models used to produce market risk metric forecasts, it is important to ensure that even models that produce very few breaches are not inefficient from an excess reserve perspective. Hence, to assess the efficiency of the forecasts produced by each model, three forecasting error measures are used, namely the mean absolute error (MAE), the root mean square error (RMSE), and the mean absolute percentage error (MAPE). In the case of zero returns, the MAPE measure would produce invalid results, and the symmetric MAPE (SMAPE) is used instead. Regardless of which of the above measures is used, a lower measure is preferred, as it represents a lower difference between return forecasts and actual returns out-of-sample.

These forecasting error measures are even more relevant where a model produces very few breaches (if any) and fails any of the various backtests. This combination often implies that the model employed results in holding excessive capital (yielding few breaches, if any) and, therefore, the model is statistically inaccurate. Hence, we use these forecasting error measures to assess model efficiency, defined to be the minimization of forecasting errors relative to the proxied profit and loss made by the equities trading desk (i.e., the minimization of excess capital reserves held).

## 4. Results
### 4.1. Value at Risk

Table 1 depicts the number of breaches observed for the various models used for the out-of-sample period spanning 15 March 1991 to 14 February 2020 to produce the 10-day 99% VaR forecasts for the S&P 500 index. The historical simulation model does not achieve the fewest breaches, ranking third and second (together with the three BNs) with three breaches achieved over the 7,286-day out-of-sample test period, when using the normal distribution and the skewed Student's t distribution as the underlying return distributions for the other models, respectively. The GARCH and EGARCH models, calibrated using the normal distribution, and the ARCH model, calibrated using the skewed Student's t distribution, experienced no breaches, whilst the largest number of breaches were experienced by the ARCH model calibrated using the normal distribution and the RiskMetrics model calibrated using the skewed Student's distribution. While the historical simulation did not achieve the fewest breaches, it is notable that it only achieved three breaches over 7,286 trading days. One of these breaches was, unsurprisingly, on 29 September 2008, indicating the model's failure to adequately react to changing market conditions experienced leading up to the 2008 global financial crisis, although this finding is not unique to the historical simulation model.

Table 1: Number of 10-day 99% Value at Risk Breaches

| Traditional Models | Normal Distribution | Skewed Student's t Distribution |
|---|---|---|
| *ARCH(1)* | 21 | 0 |
| *GARCH(1,1)* | 0 | 4 |
| *EGARCH(1,1)* | 0 | 6 |
| *RiskMetrics* | 2 | 21 |
| *Historical Simulation* | 3 | |
| *Delta-Normal* | 5 | |
| **BN Models** | | |
| *MMHC* | 3 | |
| *PC (Stable)* | 3 | |
| *SI-HITON-PC* | 3 | |

Note: This table reports the number of breaches experienced for the various 10-day value at risk (VaR) forecasting models at the 99% confidence level using either the normal distribution or the skewed Student's t distribution as the underlying return distribution (where applicable) and the daily logged returns of the Standard & Poor's (S&P) 500 index from 15 March 1991 to 14 February 2020. A VaR breach was recorded where the loss incurred on the S&P 500 index (as measured by its daily logged return) exceeded the forecasted VaR figure obtained via one of the models detailed in this table. The total number of VaR forecasts for the study period was 7,286 per model. Note that the historical simulation and the delta-normal models are distribution agnostic, meaning that the number of breaches experienced does not vary with the use of either distribution as the underlying return distribution. The Bayesian network (BN) learning algorithms considered were the max-min hill-climbing (MMHC) algorithm, the Peter and Clark Stable (PC (Stable)) algorithm, and the semi-interleaved HITON parents and children (SI-HITON-PC) algorithm.

The low number of breaches achieved by each of the traditional models tested to produce 10-day 99% VaR forecasts, relative to the much longer out-of-sample test period, is supported by the literature. McAleer and da Veiga (2008) find that achieving few breaches is preferable to achieving no breaches, as a bank would prefer using an inaccurate model that produces few breaches and incur the additional penalty by a higher penalty variable to using an accurate model that leads to higher capital requirements. The relatively lower number of breaches achieved by the historical simulation model is indicative of its poor performance, as highlighted recently by Taylor (2020). Moreover, this result highlights the historical simulation's lack of timely volatility-updating capabilities, leading to smoother forecasts relative to other models, as highlighted by Daníelsson (2002).

In the case of the BN models, breaches were all encountered on the same days as the historical simulation model, namely 27 October 1997, 31 August 1998, and 29 September 2008. The low numbers of breaches experienced by the BNs suggest that the results of the BNs are in line with those experienced by banks using traditional models to produce 10-day 99% VaR forecasts. Thus, should banks opt to employ BN models to calculate market risk, the number of breaches observed for each of the BNs would be in line with their preferences for models that produce a low non-zero number of breaches (McAleer & da Veiga, 2008).

All models, traditional and BN, pass the BCBS's traffic light test with a 'Green zone' outcome, indicating that the models tested and their resulting forecasts are acceptable based on this simple regulatory backtest. Given the few breaches achieved by each of the models relative to the number of trading days in the out-of-sample period, even by the worst performing models, it is unsurprising that the BCBS's traffic light test results in consistent 'Green zone' outcomes.

When applying Kupiec's proportion of failure test, given the 99% confidence level of the backtest and the 7,286-trading-day out-of-sample test period, the expected number of breaches for each of the models was between 50 and 94 breaches, inclusive. Since even the models producing the highest number of breaches only produced 21 breaches, all models are statistically inaccurate at the 1% significance level. As a final backtest of 10-day 99% VaR breaches, we tested the independence of breaches using Christoffersen's test for independence. With so few breaches over a relatively much longer out-of-sample test period, the few breaches observed were all deemed independent of each other at the 1% significance level.

As the historical simulation model is widely used by banks, we focus on analyzing its performance as a benchmark and its backtesting results are consistent with the literature. For example, the model conforms to regulatory requirements such as the BCBS's traffic light test, as highlighted by Sharma (2012). Moreover, supporting Sharma's (2012) findings, we find that the model's performance is mixed when using hypothesis testing using Kupiec's and Christoffersen's backtests. The few breaches exhibited also indicate the conservatism of the historical simulation model, echoing the findings of Berkowitz and O'Brien (2002).

Table 2, below, contains the MAE, the RMSE, and the MAPE values for the various 10-day 99% VaR models for the BN models and the traditional models using, where applicable, the normal and skewed Student's t distribution. Overall, the model that performed best (i.e., scoring the lowest measures) was the EGARCH model using the normal distribution across all three measures. This finding supports the findings of studies such as that of O'Brien and Szerszeń (2017), suggesting that the use of autoregressive models may produce more accurate 10-day 99% VaR forecasts. This may be explained by the tendency of autoregressive models to better capture volatility clustering and leptokurtosis (Angelidis, Benos, & Degiannakis, 2004), even when the normal distribution is used as the underlying statistical distribution, although the level of leptokurtosis induced by the autoregressive model may not capture that present in the return data (Angelidis, et al., 2004). Interestingly, the worst performing model was the ARCH model using the normal distribution, across all three measures, while the historical simulation model is the ninth-least accurate forecasting model tested, further highlighting the model's low updating abilities and its tendency to produce conservative forecasts, as highlighted by Berkowitz and O'Brien (2002).

Examining the differences in forecast accuracy due to the use of the skewed Student's t distribution as opposed to the normal distribution for the autoregressive and RiskMetrics models, the GARCH, the EGARCH, and the RiskMetrics models' forecasting accuracy deteriorated when the skewed Student's t distribution was used to produce forecasts as opposed to the normal distribution. On the other hand, the ARCH model's forecasting accuracy improved when calibrated using the skewed Student's t distribution as the underlying return distribution relative to the forecasts achieved when using the normal distribution instead.

Focusing on the three BN models used to produce the 10-day 99% VaR forecasts, they achieved results similar to the historical simulation model. This is expected given that the 10-day 99% VaR forecasts were calculated using only one BN-forecasted return, while the (much larger) remainder was made up of the actual returns observed. Hence, on each day, the differences between the 10-day 99% VaR forecasts would be minor, and the results contained in Table 2 support this.

Since the differences in the forecasting error measures' values come down to the quality of the forecasts made by the BNs, Table 2 highlights that the BN algorithm producing the worst 10-day 99% VaR forecasts (i.e., the one with the highest value for each of the three forecasting error measures) used the MMHC algorithm. On the other hand, the BNs using the PC (Stable) algorithm and the SI-HITON-PC algorithm scored equally when using the MAE and the RMSE. However, the table shows that the BN scoring the lowest MAPE value (and, therefore, ranks as the most accurate) is the BN using the SI-HITON-PC algorithm.

Table 2: Forecasting Error Measures for the 10-day 99% Value at Risk Forecasts

|  | Normal Distribution | | | Skewed Student's t Distribution | | |
|---|---|---|---|---|---|---|
| **Traditional Models** | MAE | RMSE | MAPE | MAE | RMSE | MAPE |
| *ARCH(1)* | 0.1115 | 0.1378 | 113.092% | 0.0822 | 0.0876 | 80.729% |
| *GARCH(1,1)* | 0.0715 | 0.0814 | 60.362% | 0.0728 | 0.0839 | 63.537% |
| *EGARCH(1,1)* | 0.0694 | 0.0776 | 59.719% | 0.0713 | 0.0812 | 63.338% |
| *RiskMetrics* | 0.0698 | 0.0813 | 59.793% | 0.0708 | 0.0826 | 62.8939% |
| **Distribution Agnostic** | MAE | RMSE | MAPE | | | |
| *Historical Simulation* | 0.0944 | 0.1003 | 98.135% | | | |
| *Delta-Normal* | 0.0793 | 0.0833 | 82.474% | | | |
| **BN Models** | MAE | RMSE | MAPE | | | |
| *MMHC* | 0.0945 | 0.1004 | 98.168% | | | |
| *PC (Stable)* | 0.0944 | 0.1003 | 98.136% | | | |
| *SI-HITON-PC* | 0.0944 | 0.1003 | 98.135% | | | |

Note: This table reports the results of the three forecasting error measures, namely the mean absolute error (MAE), the root mean square error (RMSE), and the mean absolute percentage error (MAPE) for the various 10-day value at risk (VaR) models at the 99% confidence levels over the period 15 March 1991 to 14 February 2020. The results are based on the logged returns earned on the Standard & Poor's (S&P) 500 index using either the normal or the skewed Student's t distribution as the underlying return distribution (where applicable) and the measures are based on the difference between the forecasted values of the models and the actual returns achieved for each trading day. The total number of forecasts for the study period was 7,286 per model. Note that the historical simulation and the delta-normal models are distribution agnostic, meaning that the forecasted values do not vary with the use of either distribution as the underlying return distribution. The Bayesian network (BN) learning algorithms considered were the max-min hill-climbing (MMHC) algorithm, the Peter and Clark Stable (PC (Stable)) algorithm, and the semi-interleaved HITON parents and children (SI-HITON-PC) algorithm.

Across both distributions, no traditional model of those tested yielded particularly accurate forecasts, regardless of the distributional assumption used, as shown by the relatively high error measures. Hence, we conclude that the general methodology used in practice today to calculate 10-day 99% VaR forecasts is inadequate and inefficient, as shown by the relatively large forecasting error measures. The similar performance displayed by the BNs means that, while not offering a significant improvement, do not offer a significant deterioration relative to the best-performing traditional model (the EGARCH model), and the best-performing BN (using the SI-HITON-PC algorithm) scored equally well to the historical simulation model.

### 4.2. Stressed Value at Risk

The breaches observed for the various models used to forecast the 10-day 99% SVaR forecasts are summarized in Table 3. There were very few breaches (and none at all for the BN models), which is expected as this metric was calculated over a stressed period, i.e., the most severe returns were used to calibrate the models and obtain forecasts. The highest number of breaches was achieved by the EGARCH and GARCH models using the normal distribution over the 7,286-day out-of-sample period. The historical simulation's non-zero number of breaches supports McAleer and da Veiga's (2008) view that banks prefer few breaches to no breaches. The finding that the forecasts produced are generally conservative with few breaches observed support those Berkowitz and O'Brien (2002), Pérignon, et al., (2008), Berkowitz, et al., (2009), Pérignon and Smith (2010), and O'Brien and Szerszeń (2017), among others, concluding that the models employed by (US) banks to calculate 10-day 99% VaR or SVaR forecasts produce conservative forecasts.

The low numbers of breaches led to 'Green zone' outcomes for all models when applying the BCBS's traffic light test to the 10-day 99% SVaR forecasts. Hence, we conclude that the BCBS's traffic light test is not a suitable test to test the statistical accuracy of a forecasting model, but, rather, it is a test to validate that a forecasting model's observed breaches are not statistically excessive, rendering the test of little use when banks' arsenals of forecasting models produce so few breaches by design.

Table 3: Number of 10-day 99% Stressed Value at Risk Breaches

| Traditional Models | Normal Distribution | Skewed Student's t Distribution |
|---|---|---|
| *ARCH(1)* | 1 | 0 |
| *GARCH(1,1)* | 2 | 0 |
| *EGARCH(1,1)* | 2 | 0 |
| *RiskMetrics* | 0 | 0 |
| **Distribution Agnostic** | | |
| *Historical Simulation* | 0 | |
| *Delta-Normal* | 0 | |
| **BN Models** | | |
| *MMHC* | 0 | |
| *PC (Stable)* | 0 | |
| *SI-HITON-PC* | 0 | |

Note: This table reports the number of breaches experienced for the various 10-day stressed value at risk (SVaR) forecasting models at the 99% confidence level using either the normal distribution or the skewed Student's t distribution as the underlying return distribution (where applicable) using the daily logged returns of the Standard & Poor's (S&P) 500 index from 15 March 1991 to 14 February 2020. A SVaR breach was recorded where the loss incurred on the S&P 500 index (as measured by its daily logged return) exceeded the forecasted SVaR figure obtained via one of the models detailed in this table, where the SVaR figure is calculated over the most severe period preceding the return's date, i.e., over a stressed period. The total number of SVaR forecasts for the study period was 7,286 per model. Note that the historical simulation and the delta-normal models are distribution agnostic, meaning that the number of breaches experienced does not vary with the use of either

distribution as the underlying return distribution. The Bayesian network (BN) learning algorithms considered were the max-min hill-climbing (MMHC) algorithm, the Peter and Clark Stable (PC (Stable)) algorithm, and the semi-interleaved HITON parents and children (SI-HITON-PC) algorithm.

The number of breaches expected for the 7,286-trading-day out-of-sample period is the same as for the VaR forecasts, being the inclusive range from 50 to 94 breaches, at the 99% confidence level. Given the few breaches observed for each of the SVaR models, all were deemed statistically inaccurate from a forecasting adequacy perspective at the 99% confidence level. Similarly, the Christoffersen's test for independence concluded that the null hypothesis cannot be rejected, i.e., all models employed exhibited independent breaches, regardless of the model and the statistical distribution used. Further, since no breaches were experienced by any of the BNs employed, it is not possible to conclude that there exists any dependence between the non-existent breaches. Hence, the backtest's null hypothesis cannot be rejected at the 1% significance level, or at any significance level. We conclude that the statistical backtests available to US banks are of little use when it comes to assessing models that produce no breaches. The only useful analysis is, therefore, that related to the relative efficiency of the models in producing 10-day 99% SVaR forecasts.

Table 4 depicts the results of the three forecasting error measures employed in this study, as calculated for the various models using the normal distribution or the skewed Student's t distribution, where applicable. Overall, the model that scored the lowest (and, therefore, performed best) was the GARCH model using the normal distribution, while the worst-performing model was the RiskMetrics model using the normal distribution.

We note that improvement in rank of the ARCH model when using the skewed Student's t distribution (ranked third using all three measures) over the normal distribution (ranked fifth using MAE and MAPE and eleventh using RMSE). The model's performance has not improved much but, rather, its climb in rank is primarily driven by the deterioration in the performances of the GARCH and the EGARCH models when using the skewed Student's t distribution instead of the normal distribution. Specifically, the two models' performances deteriorated by between 40% and 65% across the three forecasting error measures, indicating that the fit of the skewed Student's t distribution is inferior to that of the normal distribution when used to produce 10-day 99% SVaR forecasts. Hence, while we expected the skewed Student's t distribution to fit the return distribution better in general, its fit in the tail is worse relative to the fit of the normal distribution as the underlying returns distribution.

Table 4: Forecasting Error Measures for the 10-day 99% Stressed Value at Risk Forecasts

|  | Normal Distribution | | | Skewed Student's t Distribution | | |
|---|---|---|---|---|---|---|
| **Traditional Models** | MAE | RMSE | MAPE | MAE | RMSE | MAPE |
| *ARCH(1)* | 0.1270 | 0.1606 | 125.562% | 0.1117 | 0.1154 | 117.202% |
| *GARCH(1,1)* | 0.0931 | 0.0965 | 97.283% | 0.1400 | 0.1449 | 148.590% |
| *EGARCH(1,1)* | 0.0937 | 0.0977 | 97.941% | 0.1524 | 0.1564 | 159.356% |
| *RiskMetrics* | 0.3746 | 0.3929 | 394.864% | 0.1789 | 0.1922 | 190.990% |
| **Distribution Agnostic** | MAE | RMSE | MAPE | | | |
| *Historical Simulation* | 0.1482 | 0.1544 | 156.973% | | | |
| *Delta-Normal* | 0.1149 | 0.1172 | 120.334% | | | |
| **BN Models** | MAE | RMSE | MAPE | | | |
| *MMHC* | 0.1465 | 0.1523 | 154.661% | | | |
| *PC (Stable)* | 0.1482 | 0.1544 | 156.974% | | | |
| *SI-HITON-PC* | 0.1482 | 0.1544 | 156.973% | | | |

Note: This table reports the results of the four forecasting error measures, namely the mean absolute error (MAE), the root mean square error (RMSE), and the mean absolute percentage error (MAPE) for the various 10-day stressed value at risk (SVaR) models at the 99% confidence levels over the period 15 March 1991 to 14 February 2020, where the SVaR figure is calculated over the most severe period preceding the return's date, i.e., over a stressed period. The results are based on the logged returns earned on the Standard & Poor's (S&P) 500 index using either the normal or the skewed Student's t distribution as the underlying return distribution (where applicable) and the measures are based on the difference between the forecasted values of the models and the actual returns achieved for each trading day. The total number of forecasts for the study period was 7,286 per model. Note that the historical simulation and the delta-normal models are distribution agnostic, meaning that the forecasted values do not vary with the use of either distribution as the underlying return distribution. The Bayesian network (BN) learning algorithms considered were the max-min hill-climbing (MMHC) algorithm, the Peter and Clark Stable (PC (Stable)) algorithm, and the semi-interleaved HITON parents and children (SI-HITON-PC) algorithm.

Once again, it is clear that the historical simulation model, which is often used by banks as the go-to model when forecasting VaR, performed inadequately when producing 10-day 99% SVaR forecasts. The model ranked as the fourth- (using the MAE) and fifth-least accurate model (using the RMSE and the MAPE) when producing 10-day 99% SVaR forecasts.

Table 4 highlights that the various traditional models employed are not capital efficient, regardless of which error measure or distribution is used, suggesting highly conservative practices captured by the high forecasting error values. Hence, we conclude that the general methodology used in practice today to produce 10-day 99% SVaR forecasts, as is the case for 10-day 99% VaR forecasts, is inadequate and inefficient.

Turning to the BN models used to produce the 10-day 99% SVaR forecasts, the least accurate BN algorithm, as measured by all three forecasting error measures, was the PC (Stable) algorithm. Performing marginally better on only one of the three forecasting error measures is the SI-HITON-PC algorithm (which ranked as the most accurate BN algorithm when producing 10-day 99% VaR forecasts). On the other hand, the most accurate 10-day 99% SVaR forecasts were produced using the MMHC algorithm. We note that the forecasting error values in Table 4 show greater variability than their unstressed counterparts. This may be due to the use of a stress period or how it is calibrated in this study. This period would contain a BN-produced forecast if said forecast results in the next day's expected return to be included in the stressed period. On the next day, the previous day's return is now known, and the expected return is replaced with the actual return. This return may or may not be included in the stressed period. Hence, the differences in the forecasting error measures detailed in Table 4 may be due to the discrepancies in the one-day-ahead forecasts produced by the BN learning algorithms employed in this study and whether the actual return on that day is included in the stressed period. This may be arguably more meaningful than the differences in 10-day 99% VaR forecasts, given the risk management focus of the bank in this context.

## 5. Conclusion

In this study, we compared the performances of ten traditional market risk models and three BN models when producing 10-day 99% VaR forecasts and 10-day 99% SVaR forecasts using historical return data for the S&P 500 index from 15 March 1991 to 14 February 2020, mimicking the market risk management of a US bank's equities desk.

All traditional models yielded very few breaches and were classified as 'Green zone' models using the BCBS's traffic light test for both risk metrics. Hence, we conclude that the BCBS's traffic light test is of little use when testing the accuracy of either VaR or SVaR forecasts. The few breaches experienced by all traditional models concurs with the literature stating that, for banks, few breaches are preferred to no breaches (see, for example, McAleer and da Veiga, 2008). Further, the other statistical backtests used to evaluate the performances of the various traditional models when producing VaR and SVaR forecasts (Kupiec and Christoffersen backtests) yielded little additional insight. This concurs with studies such as those of Berkowitz and O'Brien (2002), Pérignon, et al., (2008), Berkowitz, et al., (2009), Pérignon and Smith (2010), and O'Brien and Szerszeń (2017), among others, which conclude that the models employed by (US) banks to calculate 10-day 99% VaR forecasts produce conservative forecasts. We extend this finding to 10-day 99% SVaR in this study.

Amongst traditional models, the EGARCH model using the normal distribution as the underlying return distribution produced the most accurate 10-day 99% VaR forecasts, while the GARCH model, also using the normal distribution, does the same for the 10-day 99% SVaR forecasts. This finding echoes the findings of studies such as that of O'Brien and Szerszeń (2017) that suggest that autoregressive models may produce more accurate 10-day 99% VaR forecasts. This may be due to these models' greater abilities to capture volatility clustering and leptokurtosis (Angelidis, et al., 2004), even when the normal distribution is used as the underlying statistical distribution. Given that US banks mostly use the historical simulation model (Pérignon & Smith, 2010), it is notable that it ranks the least or second least accurate model across the traditional models considered for each market risk metric.

Using either VaR or SVaR, the forecasts produced using the normal distribution proved to be more accurate than those produced using the skewed Student's t distribution as the underlying return distribution. This result highlights that, despite some studies suggesting that distributions with higher skew relative to the normal distribution may produce better results, such as those of Giot and Laurent (2003), Nieto and Ruiz (2016), and Lambert and Laurent (2001), this finding may not be generalized to the tail as well. While the literature indicates that the distribution itself may fit the return data better, due to the different values of skewness and kurtosis relative to the normal distribution, it is the tail that is of interest when calculating tail-based market risk metrics. This suggests that the tail fit of the skewed Student's t distribution is poorer than that of the normal distribution.

The three BNs tested yielded very few breaches for VaR, and none for SVaR, again resulting in all forecasts being classified as 'Green zone' models using the BCBS's traffic light test. Hence, we conclude that the introduction and use of BNs in producing these market risk metrics do not render the resulting forecasts and models impermissible by the regulatory test dictated by the BCBS. The SI-HITON-PC algorithm produced the most accurate forecasts for 10-day 99% VaR forecasts, while the MMHC algorithm produced the most accurate 10-day 99% SVaR forecasts.

Importantly, the use of BNs to generate market risk metrics produced results comparable to those using the traditional market risk models. Hence, this first attempt at introducing BNs to market risk management yielded reasonable results, even though they do not suggest that banks should discard their traditional models in favor of BNs. However, we leave room for future research to further refine our methodology with the ambition of potentially enhancing financial risk management practices. These BNs' results, relative to the historical simulation model, say, make sense when considering that the one-day-ahead forecast of the closing value of the S&P 500 index and, by extension, the next trading day's expected return, produced by the BN only constitutes one part of the 1,264-part calibration period, 1,263 of which are the historical returns observed in the past. Hence, even though a BN incorporates a forward-looking methodology, which should, in theory, yield more accurate market risk metrics, the equal weighting of this forecast relative to the rest of the return distribution essentially has minimal impact.

The results indicate that autoregressive models incorporating forward-looking forecasts perform better in the market risk setting than those that use backwards-looking data exclusively, such as the historical simulation model, for example, or the BN models, which used equal weighting for both historical returns and the one-day-ahead forecast. Thus, there is a strong case for incorporating weighted forward-looking forecasts beyond simply the volatility of returns and to examine the performances of models that incorporate the forecasts of the returns themselves. BNs have been increasingly used in financial settings, with a recent relevant application being that of Apps (2020). Hence, the incorporation of a weighted forward-looking methodology to the market risk field can be achieved using BNs to forecast the closing values of the S&P 500 index, which can then be weighted and used to produce market risk metrics.

Future research may, therefore, focus on refining our methodology, perhaps via an attempt to weight the forecasts made by the BN when calculating either VaR or SVaR, or expand on it by using additional algorithms and applying it to different trading desks. Moreover, the relatively recent move by the BCBS to using ES and its stressed counterpart requires future research to assess the viability of BNs in producing these market risk metrics, too.

# Appendix A: Variables used to Train the Bayesian Networks

Table 5: Variables used to train the various Bayesian networks

| Variable Name | Classification |
| --- | --- |
| *S&P 500 index (closing value)* | Target variable |
| *Australian Dollar/US Dollar* | Currency Exchange Rate |
| *Bloomberg Commodities Index* | Financial |
| *Bloomberg US Treasury Total Return Index* | Financial |
| *Canadian Dollar/US Dollar* | Currency Exchange Rate |
| *Canola Price* | Commodity |
| *Corn Price* | Commodity |
| *Euro/US Dollar* | Currency Exchange Rate |
| *Federal Reserve Lending Rate* | Economic |
| *Financial Times Stock Exchange 100 Index* | Financial |
| *Gold Price* | Commodity |
| *Great British Pound/US Dollar* | Currency Exchange Rate |
| *Japanese Yen/US Dollar* | Currency Exchange Rate |
| *Nikkei Index* | Financial |
| *Oats Price* | Commodity |
| *Russell 1000 Index* | Financial |
| *Russell 2000 Index* | Financial |
| *Russell 3000 Index* | Financial |
| *Silver Price* | Commodity |
| *Soybean Meal Price* | Commodity |
| *Soybean Oil Price* | Commodity |
| *Soybean Price* | Commodity |
| *S&P Australian Stock Exchange Index* | Financial |
| *S&P Toronto Stock Exchange Index* | Financial |
| *Three-month US London Interbank Offer Rate* | Financial |
| *Topix Index* | Financial |
| *US Consumer Price Index* | Economic |
| *US Corporate Aaa-rated Ten-year Spread* | Financial |
| *US Corporate Bonds Index* | Financial |
| *US Disposable Income Growth Index* | Economic |
| *US Full Employment* | Economic |
| *US Jobless Claims* | Economic |
| *US M1 Money Supply* | Economic |
| *US M2 Money Supply* | Economic |
| *US s Manufacturing Index* | Economic |
| *US Manufacturing Tendency Index* | Economic |
| *US Non-Farm Payroll* | Economic |
| *US Part Time Employment* | Economic |
| *US Policy Uncertainty Index* | Political |
| *West Texas Intermediate Price* | Commodity |
| *Wheat Price* | Commodity |

Note: This table lists the variables used to train the Bayesian networks (BNs) using the various learning algorithms in this study. The data for the variables were obtained from the Bloomberg database over the period 15 March 1991 to 14 February 2020, as well as for a calibration period and a training period preceding this start date. The data relate to variables from the United States (US) as identified (from the literature or otherwise) to causally relate to the target variable, the Standard & Poor's (S&P) 500 index.